\documentclass[10pt, paper=a4, UKenglish]{article}
\usepackage{graphicx}
\def\Title#1{\begin{center} {\Large #1 } \end{center}}
\def\Author#1{\begin{center}{ \sc #1} \end{center}}
\def\Address#1{\begin{center}{ \it #1} \end{center}}

\newcommand\pubblock{\rightline{\begin{tabular}{l} Proceedings of the CTD 2023\\ \pubnumber\\
         \pubdate  \end{tabular}}}

\newenvironment{Abstract}{\begin{quotation} \begin{center} 
             \large ABSTRACT \end{center}\bigskip 
      \begin{center}\begin{large}}{\end{large}\end{center} \end{quotation}}

\newenvironment{Presented}{\begin{quotation} \begin{center} 
             PRESENTED AT\end{center}\bigskip 
      \begin{center}\begin{large}}{\end{large}\end{center} \end{quotation}}

\def\Acknowledgements{\bigskip  \bigskip \begin{center} \begin{large}
      \bf ACKNOWLEDGEMENTS \end{large}\end{center}}

\def\beq{\begin{equation}}
\def\eeq#1{\label{#1}\end{equation}}
\def\eeqn{\end{equation}}

\def\beqa{\begin{eqnarray}}
\def\eeqa#1{\label{#1}\end{eqnarray}}
\def\eeqan{\end{eqnarray}}

\let\bar=\overbar

\def\Dslash{\not{\hbox{\kern-4pt $D$}}}
\def\dslash{\not{\hbox{\kern-2pt $\del$}}}

\def\msb{{\bar{\ssstyle M \kern -1pt S}}}

\usepackage{color}
\usepackage{lineno}
\usepackage{subfig}
\usepackage{hyperref}

\usepackage{xspace}
\usepackage[numbers,sort&compress]{natbib}
\graphicspath{{figures/}}

\newcommand{\triton}{Triton\texttrademark\xspace}

\newcommand\pubnumber{PROC-CTD2023-56}

\newcommand\pubdate{\today}

\newcommand{\conference}{Connecting the Dots Workshop (CTD 2023)\\
October 10-13, 2023}

\usepackage{fancyhdr}
\definecolor{mygrey}{RGB}{105,105,105}
\begin{document}


\large
\begin{titlepage}
\pubblock

\vfill
\Title{Graph Neural Network-based Tracking as a Service}
\vfill

\Author{Haoran Zhao~$^1$, Andrew Naylor~$^3$, Shih-Chieh Hsu~$^1$, Paolo Calafiura~$^2$,  Steven Farrell~$^3$, Yongbing Feng~$^4$, Philip Coleman Harris~$^5$, Elham E Khoda~$^1$,  William Patrick Mccormack~$^5$,  Dylan Sheldon Rankin~$^6$, Xiangyang Ju~$^2$}
\Address{
$^1$ University of Washington, Seattle, WA 98195, USA \\
$^2$ Lawrence Berkeley National Laboratory, Berkeley, CA 94720, USA \\
$^3$ National Energy Research Scientific Computing Center, Berkeley, CA 94720, USA \\
$^4$ Fermi National Accelerator Laboratory, Batavia, IL 60510, USA \\
$^5$ Massachusetts Institute of Technology, Cambridge, MA 02139, USA \\
$^6$ University of Pennsylvania, Philadelphia, PA 19104, USA
}
\vfill

\begin{Abstract}
Recent studies have shown promising results for track finding in dense environments using Graph Neural Network (GNN)-based algorithms. However, GNN-based track finding is computationally slow on CPUs, necessitating the use of coprocessors to accelerate the inference time. Additionally, the large input graph size demands a large device memory for efficient computation, a requirement not met by all computing facilities  used for particle physics experiments, particularly those lacking advanced GPUs. Furthermore, deploying the GNN-based track finding algorithm in a production environment requires the installation of all dependent software packages, exclusively utlized by this algorithm. These computing challenges must be addressed for the successful implementation of GNN-based track finding algorithm into production settings. In response, we introduce a ``GNN-based tracking as a service'' approach, incorporating a custom backend within the \triton inference server to facilitate GNN-based tracking. This paper presents the performance of this approach using the Perlmutter supercomputer at NERSC.

\end{Abstract}

\vfill

\begin{Presented}
\conference
\end{Presented}
\vfill
\end{titlepage}
\def\thefootnote{\fnsymbol{footnote}}
\setcounter{footnote}{0}
%

\normalsize 


\section{Introduction}
\label{sec:intro}
The ExaTrkX~\cite{ExaTrkX:2021abe} and the L2IT group~\cite{Biscarat:2021dlj} pioneered a Graph Neural Network (GNN)-based pipeline for particle track finding at the High-Luminosity Large Hadron Collider (HL-LHC) era. Recent studies~\cite{Caillou:2815578,Caillou:2871986} showed that the pipeline achieves high tracking reconstruction efficiency in the $t\bar{t}$ events with 200 additional secondary proton-proton collision events (i.e. \textit{pileup}) using the ATLAS next generation Inner detector (ITk)~\cite{ATLAS-TDR-25,ATLAS-TDR-30}. Furthermore, the track candidates from the pipeline, when integrated with the ATLAS global $\chi^2$ track fitting algorithm~\cite{ATLAS:2020ixw}, showed excellent track parameter resolutions, underlining the potential for production use. However, several computational challenges need to be addressed.

First of all, the pipeline is computationally slow with CPUs~\cite{Lazar:2022ixi}, necessitating the use of co-processors like GPUs to accelerate the inference. The substantial graph sizes involved, featuring approximately 300k nodes and 1 million edges, demand large device memory, a requirement not universally met across WLCG grid sites used by ATLAS.

Secondly, the pipeline's dependency on an extensive list of external packages presents a challenge. Integrating these packages into existing production frameworks like Athena complicates package management and sometimes leads to incompatibilities, especially with CUDA-related packages. This challenge is not unique to GNN-based tracking but extends to all machine learning (ML) inference applications. 

With the emerging of diverse coprocessors like TPUs, IPUs, FPGAs, aimed at accelerating ML inference, it is becoming impractical for production frameworks to support all types of co-processors.

\begin{figure}[htb]
    \centering
    \includegraphics[width=0.9\textwidth]{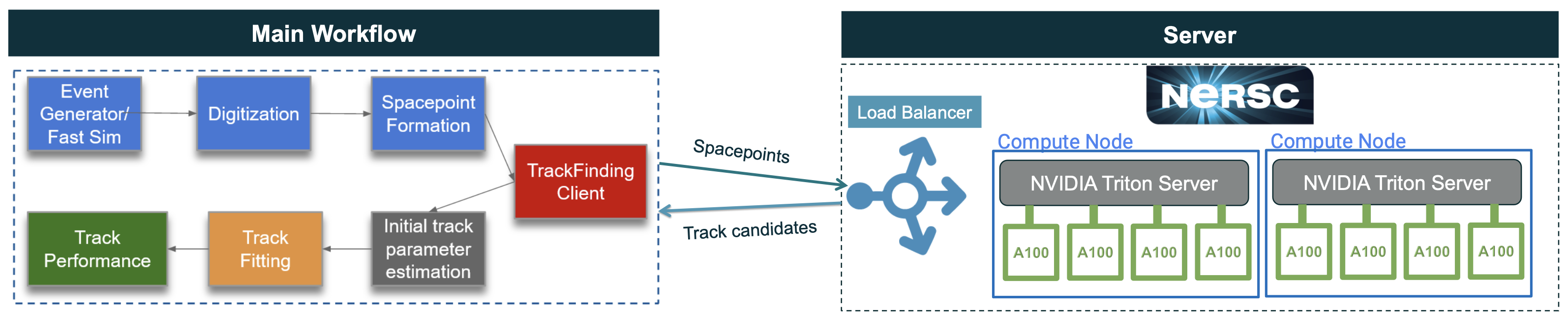}
    \caption{Workflow of the ExaTrkX pipeline.}
    \label{fig:workflow}
\end{figure}

To overcome these challenges, we propose a ``GNN-based tracking as a service'' approach, as illustrated in Fig~\ref{fig:workflow}. This approach involves developing a \triton client within the production framework that sends requests to a \triton server. The server processes the requests and returns the output to the client. In this way, the client does not need to know the GNN-based tracking implementation.

Other researchers explored similar as-a-Service approach~\cite{aas:neutrino,aas:gpus,aas:fpga} for different experiments. The CMS experiment employed a service model to integrate GPUs into the miniAOD production workflow, resulting in a significant speedup compared to CPU-only workflows~\cite{CMS-DP-2023-037}. The DUNE project implemented a cloud-based GPU-accelerated inference server for faster event reconstruction in neutrino data batch jobs~\cite{aas:dune}. Gravitational experiments adopted a similar approach for ML-assisted real-time noise regression~\cite{aas:largo}. Furthermore, a toolkit for FPGAs-as-a-Service~\cite{aas:fpga-tools} is developed, showcasing the adaptability of this model across different hardware platforms.

\section{ExaTrkX as-a-Service}

\subsection{ExaTrkX}
\begin{figure}
    \centering
    \includegraphics[width=0.8\textwidth]{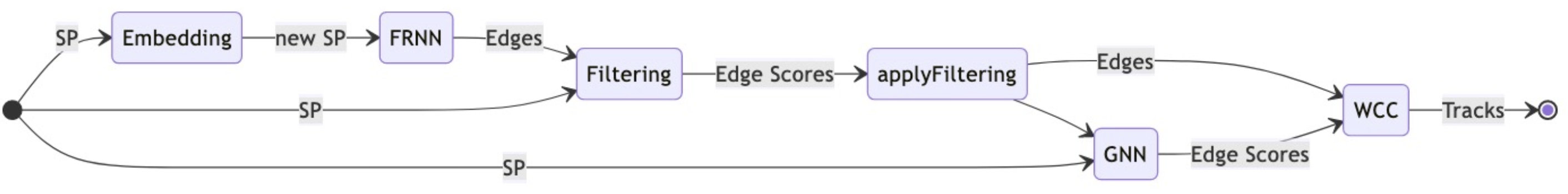}
    \caption{Triton inference ensemble backend for the ExaTrkX pipeline.}
    \label{fig:dataflow}
\end{figure}

The ExaTrkX pipeline, illustrated as a dataflow diagram in Fig.~\ref{fig:dataflow}, uses space points (\textit{SP}) as inputs and produces track candidates as outputs. It encompasses three major modules: graph construction, edge classification, and graph segmentation.  The graph construction module uses dense neural networks called Multi-Layer Perceptrons (MLPs) to encode space point raw features to a new latent space in which hits from the same track are close to each other and away from other hits. This step is called \textit{Embedding}. Then, a fixed-radius algorithm builds connections between SPs in this latent space (\textit{edges}). The number of edges after the embedding step necessitates another MLP to filter out clearly fake edges - called \textit{Filtering}. A fixed threshold applied to these filtered edge scores (referred to as applyFiltering) helps retain true edges while eliminating false ones. The edge classification step uses the interaction network~\cite{gnn:IN}. In the end, the GNN edge scores are passed to the weekly-connected-component (WCC) algorithm to form track candidates.

The ExaTrkX pipeline are implemented in \textsc{C++} for CPU-only and GPU-only~\footnote{The software can be found at https://github.com/The-ExaTrkX-Project/exatrkx-service.}. Trained \textsc{PyTorch} models are executed via the \textsc{libTorch} library. The nearest neighbour search is performed with the \textsc{FRNN} package~\cite{frnn} for GPUs and \textsc{FAISS}~\cite{faiss} for CPUs. The connected component algorithm is from the \textsc{Boost} package, with potential future improvements from the CUDA-based version in \textsc{cuGraph}.

\subsection{\triton client}

Inspired by the services for optimized network inference on coprocessors (SONIC) approach and examples from the \triton client~\cite{triton:client}, we developed \triton client utlities. The utility package helps package the input data into a remote protocol package, sends the package to the \triton server, and decodes the returned data bytes.

\subsection{Inference server}
The NVIDIA \triton Inference Server, an open-source platform, standardizes model deployment and execution, targeting fast and scalable ML in production~\cite{triton:server}. It provides a built-in metrics endpoint~\cite{triton:metrics} that publishes plain-text data in Prometheus format~\cite{prometheus} for system monitoring.

In the ExaTrkX pipeline, \textsc{Pytorch} backend is used for Embedding, Filtering, applyFiltering, and GNN; and the \textsc{Python} backend for FRNN and WCC. The dataflow dependencies are implemented with the \triton ensemble backend,orchestrating model executions based on predefined dependency schemes. 

However, the ensemble backend can be inefficient for complex workflows like ExaTrkX, often necessitating data movement between models and creating overheads. It may force users to create ad-hoc models to perform simple operations that are not supported by the existing backends. For example, the \textit{applyFiltering} model merely removes edges below a certain score threshold. Since the \textsc{Pytorch} backend does not support applying selections on its outputs, a seperate model is required for this task and data has to be moved from the \textsc{PyTorhc} model to the ad-hoc model. Moreover, its greedy scheduling can cause overheads or crashes in multi-GPU setups due to mismatched data and model locations.

An alternative and favorable solution for the ExaTrkX as-a-Service is to build a custom backend. This involves integrating the ExaTrkX \textsc{C++} implementation into the Triton Server interface and constructing a new backend library. The inputs are defined as a 2D vector (rows representing space points and columns their raw features), with outputs as a vector of integers labeling the space points.

\section{Results}
\label{sec:results}
The ExaTrkX as a Service is evaluated on the Perlmutter system at NERSC.
Each Perlmutter CPU node contains two 2x AMD EPYC 7763 (Milan) CPUs, each CPU containing 128 physics cores. Each Perlmutter GPU node contains four A100 GPUs and one AMD EPYC 7763 (Milan) CPU. 
For our tests, we configured the Triton server and client to operate within the same computing node.

Using the \triton analyze tool, we assessed various performance metrics such as model execution time, waiting time, input data processing time, and others. Detector hits were formatted into JSON for compatibility with the tool, enabling us to accurately measure the throughput of the GNN-based service.

The study uses the ExaTrkX pipeline trained for identifying large radius tracks in the Heavy Neural Lepton events as detailed in the previous study (Ref.~\cite{Wang:2022oer}).

\subsection{CPU-based Pipeline Throughput}

\begin{figure}[htb]
    \centering
    \includegraphics[width=0.6\textwidth]{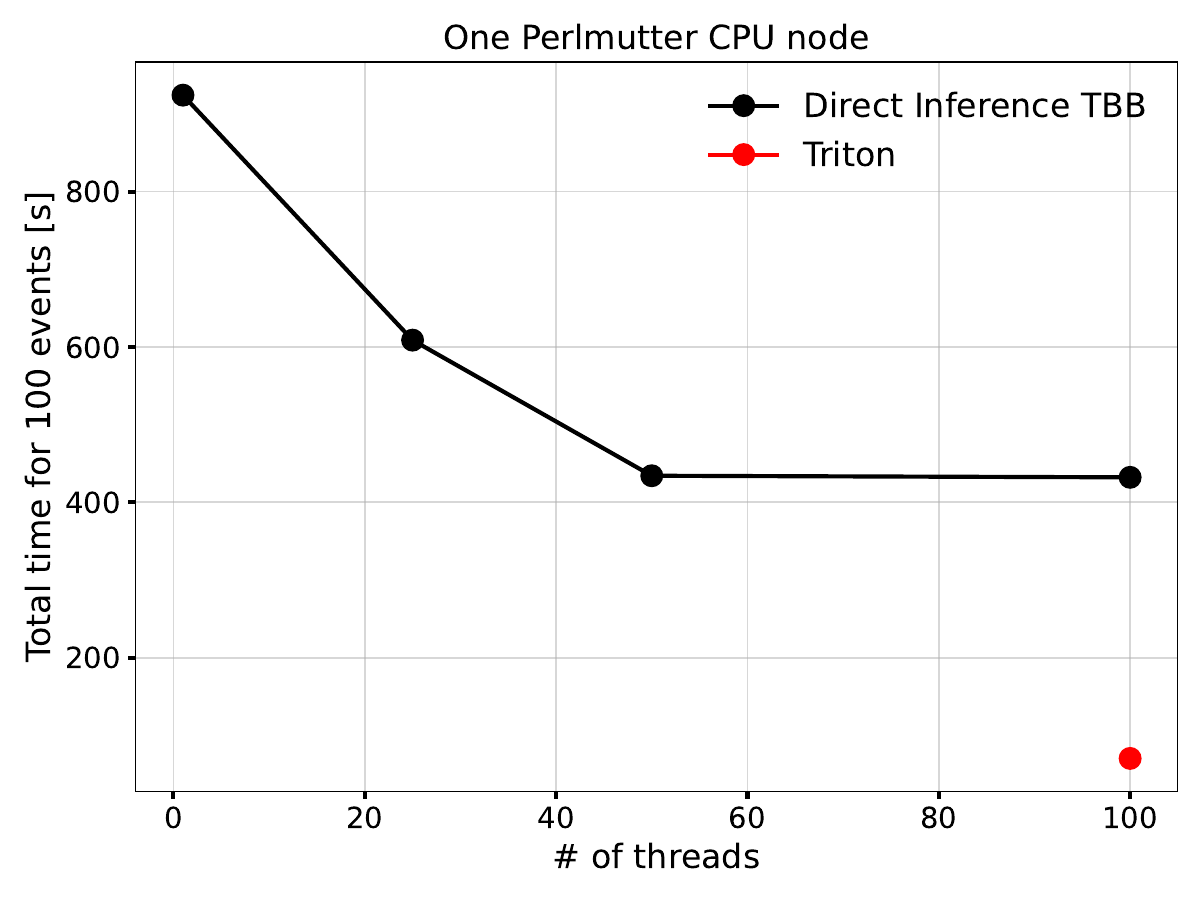}
    \caption{Inference time for 100 events as a function of the CPU threads for the direct inference and Triton.}
    \label{fig:res_cpu}
\end{figure}

As shown in Figure~\ref{fig:res_cpu}, the throughput of the CPU-based ExaTrkX pipeline varies with the number of threads. Direct CPU-based inference shows reduced inference time up to about 50 threads. Remarkably, running the CPU-based ExaTrkX as a service significantly outperforms direct running. We observed increased CPU core utilization when operating as a service. This performance improvement might be attributed to better thread management and reduced conflicts between the framework and the external library FAISS, both of which use TBB for thread management.

\subsection{GPU-based Pipeline Throughput}

\begin{figure}[htb]
    \centering
    \subfloat[]{\includegraphics[width=0.44\textwidth]{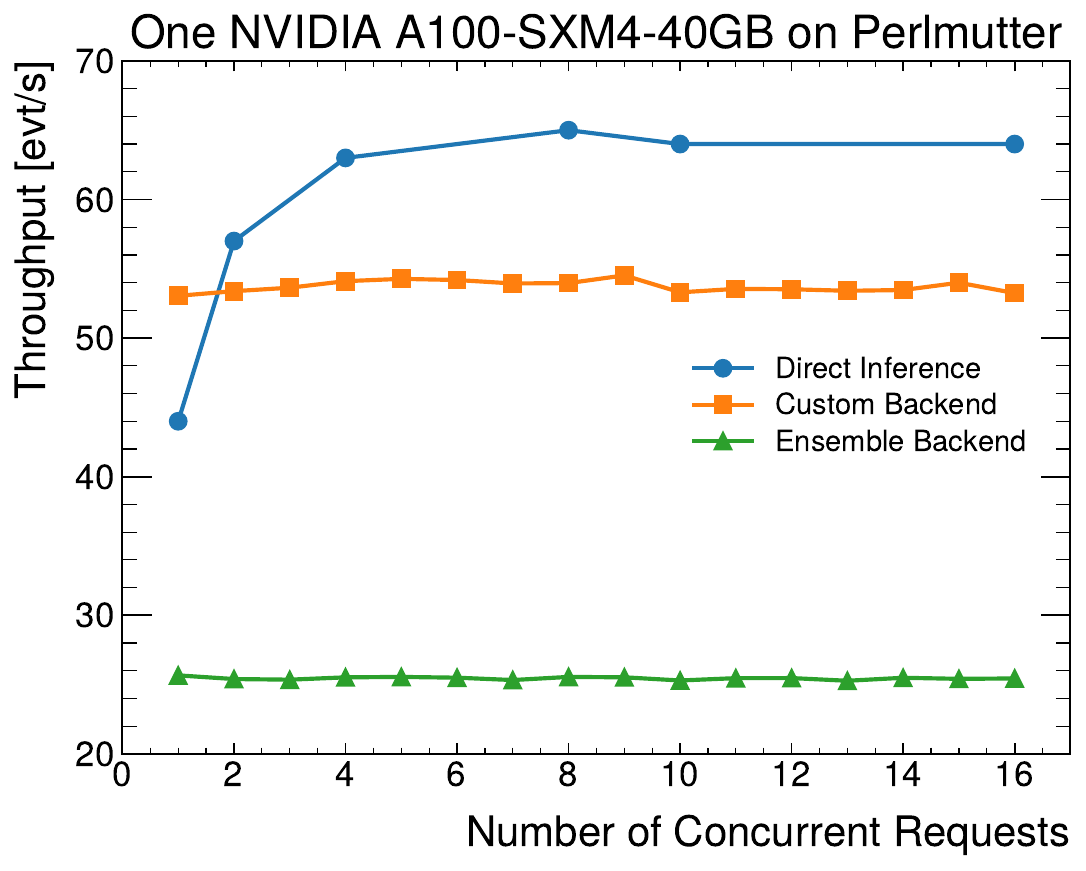}}
    \subfloat[]{\includegraphics[width=0.5\textwidth]{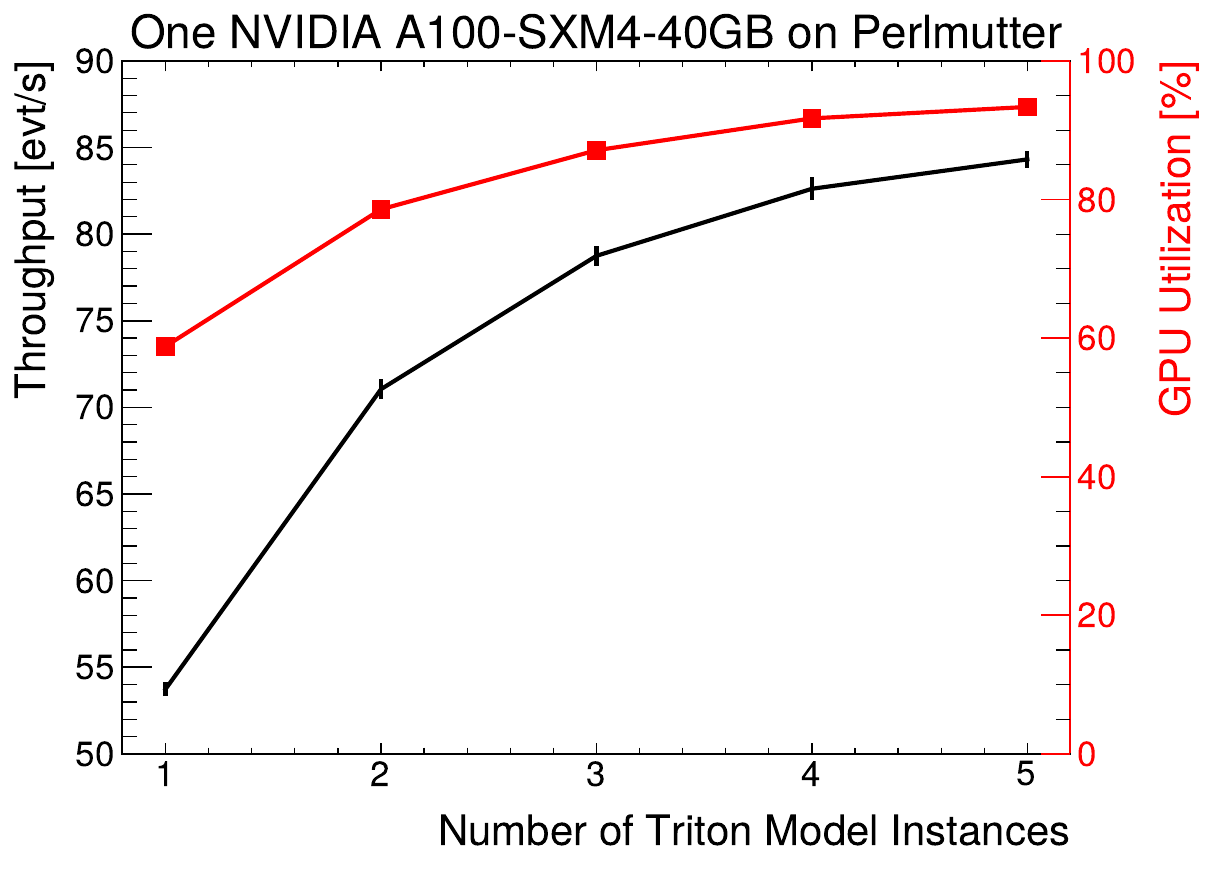}}
    \caption{(a) Throughput as a function of the number concurrent requests for direct inference (solid blue circle), Custom Backend (yellow square), and Ensemble Backend (green triangle). (b) Throughput (left $y$-axis) and GPU utilization (right red $y$-axis) as a function of the number of Triton model instances.
    }
    \label{fig:res_gpu}
\end{figure}

Figure~\ref{fig:res_gpu} shows the throughput for the GPU-based ExaTrkX pipeline. For direct GPU-based inference, an increase in threads improves throughput up to four threads, peaking at about 65 events per second. Running the GPU-based ExaTrkX as a service with the custom backend doubles the throughput compared to using the ensemble backend. Triton server processes concurrent requests sequentially. Therefore, increasing the number of concurrent requests from the client does not increase the throughput. However, creating more Triton model instances allows simultaneous handling of client requests, boosting total throughput and GPU utilization.

\section{Conclusions}
In this study, we successfully implemented the ExaTrkX pipeline in both CPU and GPU formats using C++. These implementations are compatible with production frameworks like ACTS and Athena. Our exploration of various Triton Server configurations revealed that the custom backend significantly outperforms the ensemble backend, achieving twice the speed. Moreover, we discovered that increasing the number of model instances notably enhances both throughput and GPU utilization. 

\subsection{Future Directions}

The current study used a dataset without pileup, which is substantially smaller than what is expected at the High Luminosity LHC. Our next step involves evaluating performance using a more realistic setup. This will provide insights into the scalability and efficiency of our implementation under more demanding conditions.

We aim to assess the performance of the ExaTrkX pipeline tailored for the ATLAS ITk. This evaluation will offer valuable information on the pipeline's applicability and effectiveness in a state-of-the-art experimental environment.

Building upon the work of Ref~\cite{ben:demo}, which integrates the ExaTrkX pipeline with the conventional Kalman Filter-based tracking fitting algorithm, we plan to establish this combined approach as a service. This development will potentially streamline the tracking process, offering a more comprehensive and efficient tracking solution.


\Acknowledgements
This research was supported in part by the U.S. Department of Energy’s Office of Science, Office of High Energy Physics, under Contracts No. DE-AC02-05CH11231 (CompHEP Exa.TrkX) and No. KA2102021235 (US ATLAS Operations), and by the Exascale Computing Project (17-SC-20-SC), a joint project of DOE’s Office of Science and the National Nuclear Security Administration. This research used resources from the National Energy Research Scientific Computing Center (NERSC), a U.S. Department of Energy Office of Science User Facility located at Lawrence Berkeley National Laboratory, operated under Contract No. DE-AC02-05CH11231. Zhao, Harris and Hsu are supported by National Science Foundation (NSF) grants No. 2117997. 

\bibliography{eprint}
\bibliographystyle{JHEP}

\end{document}